\documentclass{iau}
\usepackage{graphicx}

\title[ExoplANETS-A] 
{Exoplanet host-star properties: the active environment of exoplanets} 

\author[John P. Pye et al.]   
{John P. Pye$^1$, David Barrado$^2$, Rafael A. Garc\'ia$^3$, Manuel G\"udel$^4$, Jonathan Nichols$^1$, Simon Joyce$^1$, 
Nuria Hu\'elamo$^2$, Mar\'ia Morales-Calder\'on$^2$, Mauro L\'opez$^2$, Enrique Solano$^2$, 
Pierre-Olivier Lagage$^3$, Colin P. Johnstone$^4$, Allan Sacha Brun$^3$, Antoine Strugarek$^3$, J\'er\'emy Ahuir$^3$ \\
On behalf of the ExoplANETS-A Consortium }

\affiliation{$^1$Department of Physics \& Astronomy, University of Leicester, Leicester, LE1 7RH, U.K.  \\ email: {\tt pye@le.ac.uk} \\[\affilskip]
$^2$Depto. Astrof\'isica, Centro de Astrobiolog\'ia (INTA-CSIC), ESAC Campus, 
Camino Bajo del Castillo s/n, 28692, Villanueva de la Ca\~nada, Spain  \\[\affilskip]
$^3$AIM, CEA, CNRS, Universit\'e Paris-Saclay, Universit\'e Paris Diderot, Sorbonne Paris Cit\'e, F-91191 Gif-sur-Yvette, France  \\[\affilskip]
$^4$University of Vienna, Dept.\ of Astrophysics, T\"urkenschanzstr.\ 17, 1180 Vienna, Austria
}

\pubyear{2019}
\volume{345}  
\jname{Origins: from the Protosun to the First Steps of Life}
\editors{Bruce G. Elmegreen, L. Viktor T\'oth, Manuel G\"udel, eds.}
\begin{document}

\maketitle

\begin{abstract}
The primary objectives of the ExoplANETS-A project are to:
establish new knowledge on exoplanet atmospheres; 
establish new insight on influence of the host star on the planet atmosphere; 
disseminate knowledge, using online, web-based platforms.
The project, funded under the EU's Horizon-2020 programme, started in January 2018 and has a duration $\sim 3$ years.
We present an overview of the project, the activities concerning the host stars and some early results on the host stars.
\keywords{methods: data analysis, catalogs, stars: activity, stars: atmospheres, stars: coronae, (stars:) planetary systems, stars: winds, outflows, ultraviolet: stars, X-rays: stars }
\end{abstract}

\firstsection 
\section{Introduction}

Seven institutes\footnote{CEA Saclay, Paris, France; CAB-INTA, Madrid, Spain; MPIA, Heidelberg, Germany; University College London, U.K.; University of Leicester, U.K.; SRON, Utrecht, NL; Universitat Wien, Austria} 
in Europe have combined their expertise in the field of exoplanetary research to develop the European Horizon-2020 
ExoplANETS-A\footnote{{\it Exoplanet Atmosphere New Emission Transmission Spectra Analysis}; 
https://cordis.europa.eu/project/rcn/212911\_en.html ; 
The ExoplANETS-A project has received funding from the EU's Horizon-2020 programme; Grant Agreement no.~776403.} 
project under the coordination of CEA Saclay. In the framework of the project, novel data calibration and spectral extraction tools, as well as novel retrieval tools, based on 3D models of exoplanet atmospheres, will be developed to exploit archival data from space- and ground-based observatories, and produce a homogeneous and reliable characterization of the atmospheres of transiting exoplanets. Additionally, to model successfully the exoplanet atmosphere, it is necessary to have a sound knowledge of the host star. To this end, we will collect a coherent and uniform database of the relevant properties of host stars from online archives (e.g. XMM-Newton, Gaia) and publications. These exoplanet and host-star catalogues will be accompanied by computer models to assess the importance of star--planet interactions, for example the `space weather' effects of the star on its planetary system. The knowledge gained from this project will be published through peer-reviewed scientific journals and modelling tools will be publicly released. 

\begin{figure}[h]
\begin{center}
 \includegraphics[width=50mm]{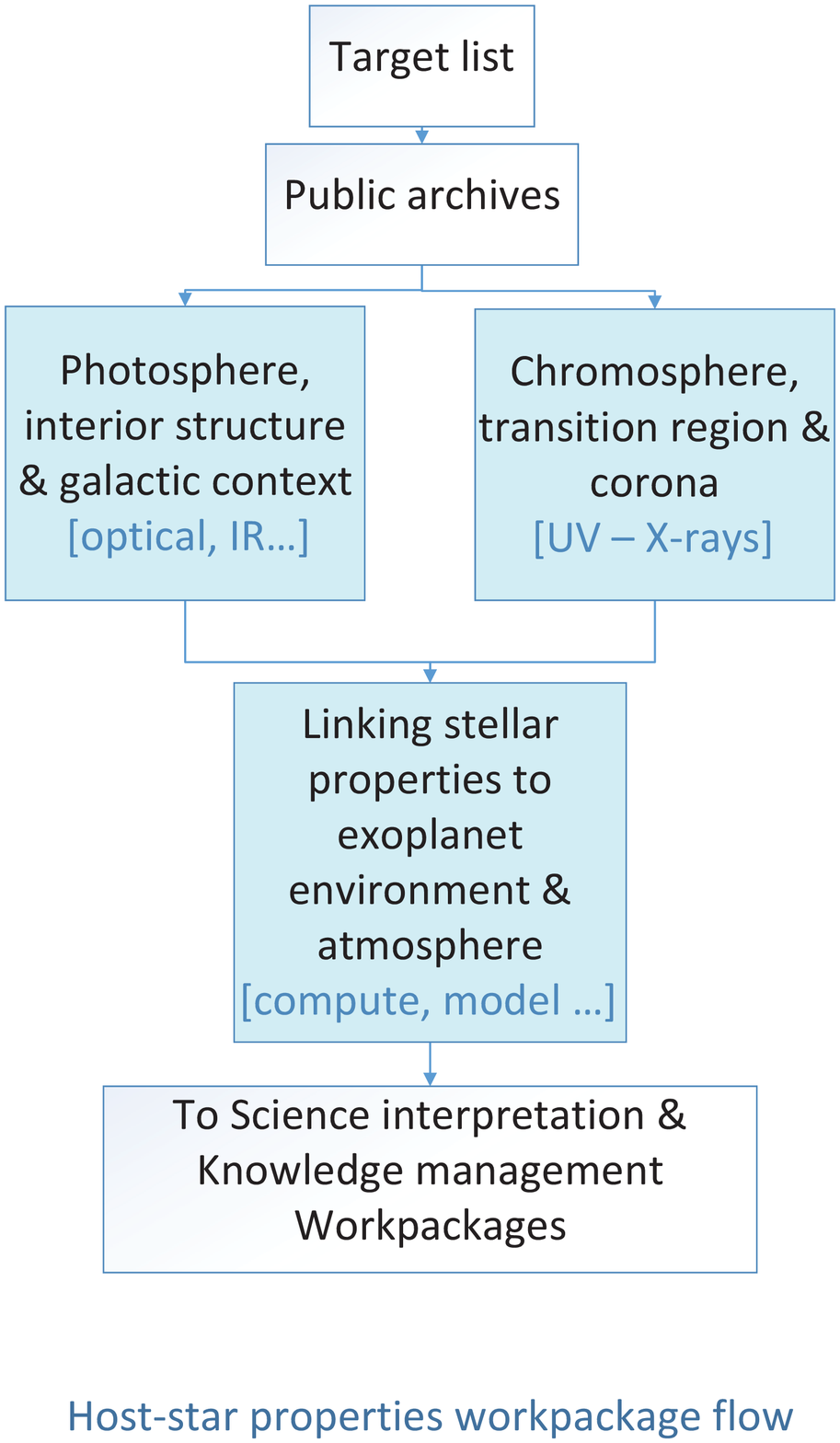}
 \includegraphics[width=80mm]{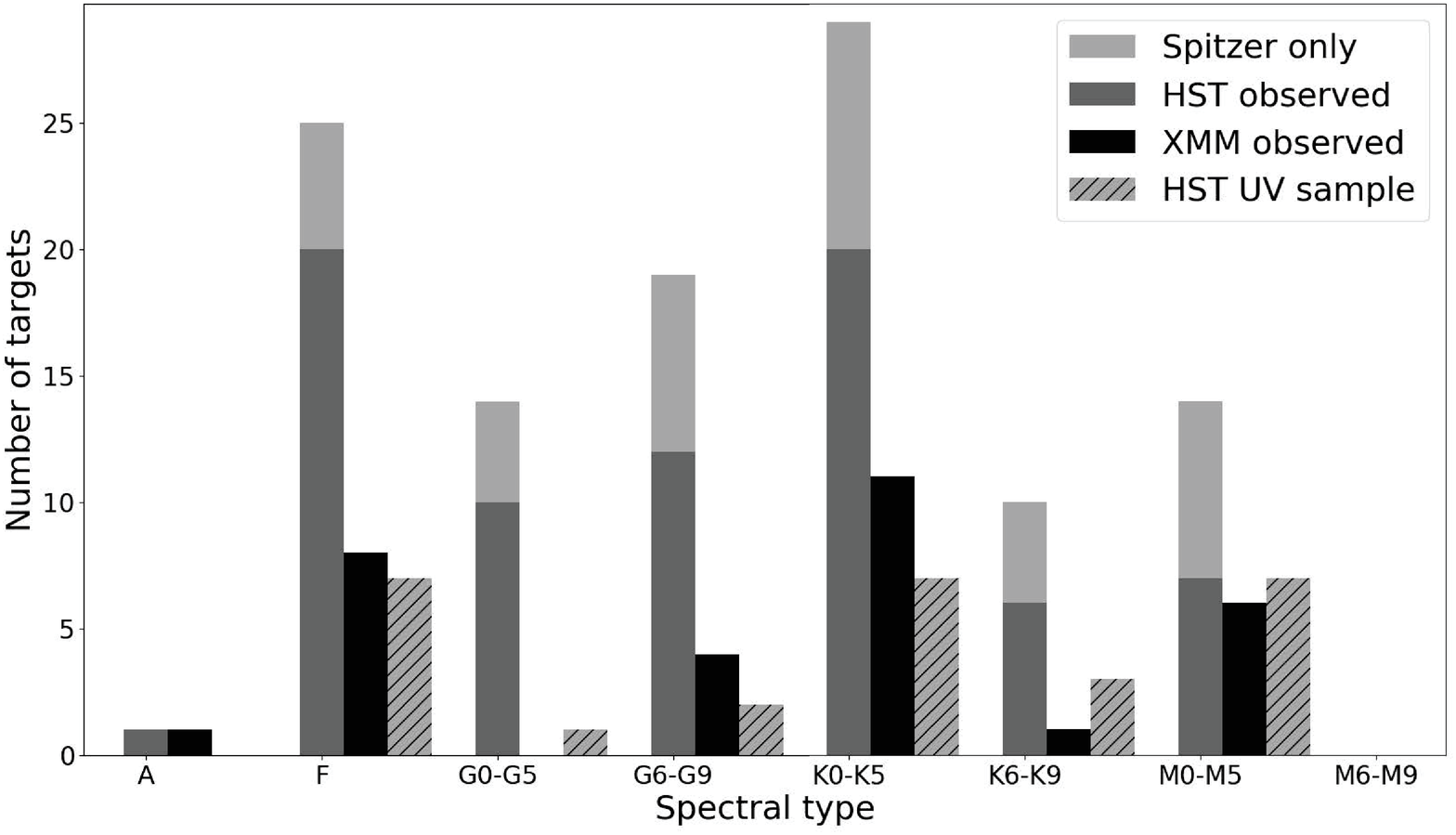}
  \caption{ {\it (a, left).} Schematic of the flow of data through the host-stars workpackage, and interfaces to the overall project. {\it (b, right).} The target sample, in terms of the distribution of host stars with spectral type, showing the breakdown by archival-observation types.}
   \label{fig_dataflow}
\end{center}
\end{figure}

The project has six work packages (WPs); the focus in this paper is on the WP `Host-star properties: the active environment of exoplanets'. Fig.\ref{fig_dataflow}(a) illustrates the flow of data through the host-stars WP, and interfaces to the overall project.

We outline the activities concerning the host stars, and present early results from the host-star investigations, including the basic stellar observational and physical properties, and indications of future observations needed to maximize the coverage of the target list of $\sim 100$ stars. We also discuss some of the modelling aspects.

\section{The host-stars activities}

In addition to compiling the observational data, we will
where necessary, e.g.\ in the EUV, interpolate, extrapolate and scale spectral and other information, to cover gaps in the observational data, in order to provide the full XUV spectral range for modelling the host-star influence on the exoplanet atmosphere (e.g.\ Nemec \etal\ 2019).
This work will be accompanied by computer models to assess the importance of star--planet interactions, for example the `space weather' effects of the star on its planetary system; We will also model the possible evolutionary scenarios for the stellar activity over the star's lifetime, in order to gain insight into the past environment of the exoplanet (e.g.\ Johnstone \etal\ 2015).

\section{The sample}

The sample of exoplanets and host stars considered by the project comprises all transiting-exoplanet systems observed by HST or Spitzer. Currently, this corresponds to 135 exoplanets, of which 85 have HST data; the associated number of stars is 113, with 76 having HST data (Fig.\ref{fig_dataflow}(b)). 
For the host stars, the primary online archival databases to be used are HST (for UV spectra), XMM-Newton (for X-ray), Gaia (for astrometric properties) and SIMBAD (for spectral types etc). The searches also include: GALEX (UV photometry); Chandra, ROSAT and other X-ray catalogues, together with results retrieved from the published literature. At X-ray wavelengths, most of the detections are from the 3XMM (DR8) serendipitous source catalogue (Rosen \etal\ 2016). 

In summary, to date (November 2018), we have the following statistics:
\begin{itemize}
	\item X-ray (3XMM-DR8 cat.): 31 stars observed, with 17 detections in the public archive, and the outcomes awaited for the others;
	\item UV photometry (GALEX, GR6 cat.): 70 stars detected (51 in the HST sample)
	\item UV spectra (HST COS and/or STIS): 26 stars observed
	\item X-ray observation \& UV photometry: 23 stars
	\item X-ray observation \& UV spectra: 20 stars
\end{itemize}
All the stars with XMM observations also have HST visible/IR spectra.

\begin{figure}[h]
\begin{center}
 \includegraphics[width=50mm]{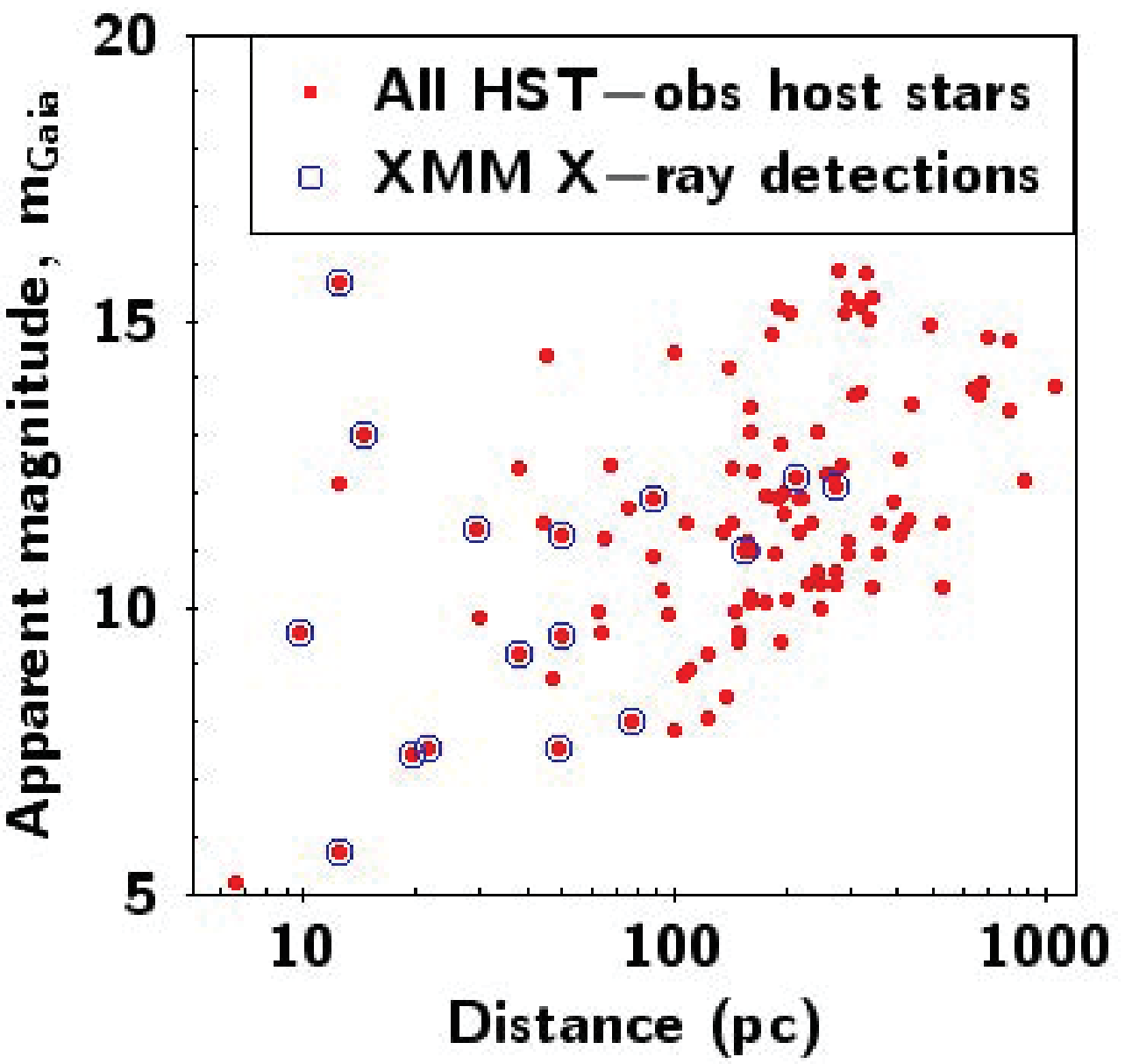}
 \includegraphics[width=40mm]{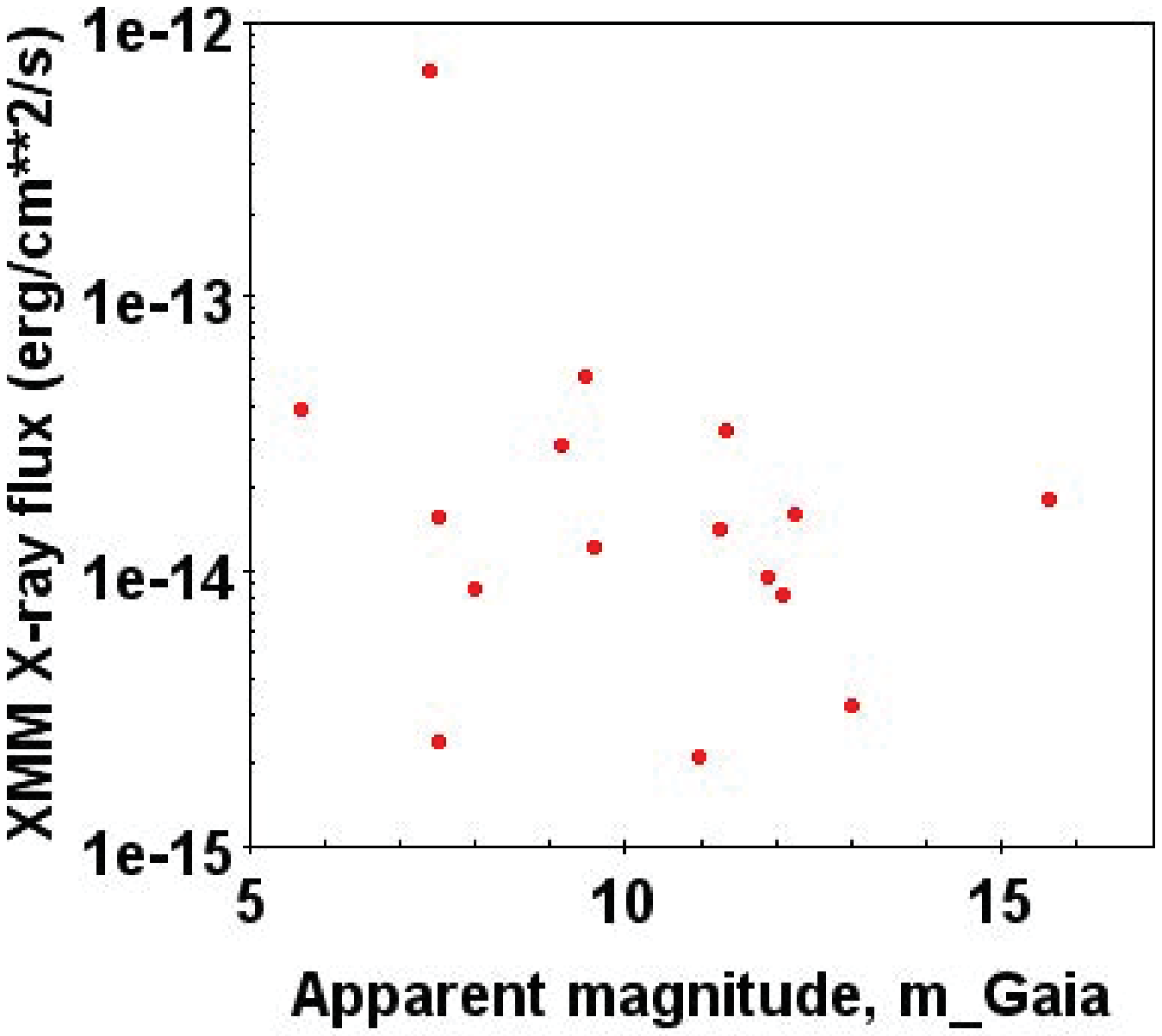}
 \includegraphics[width=40mm]{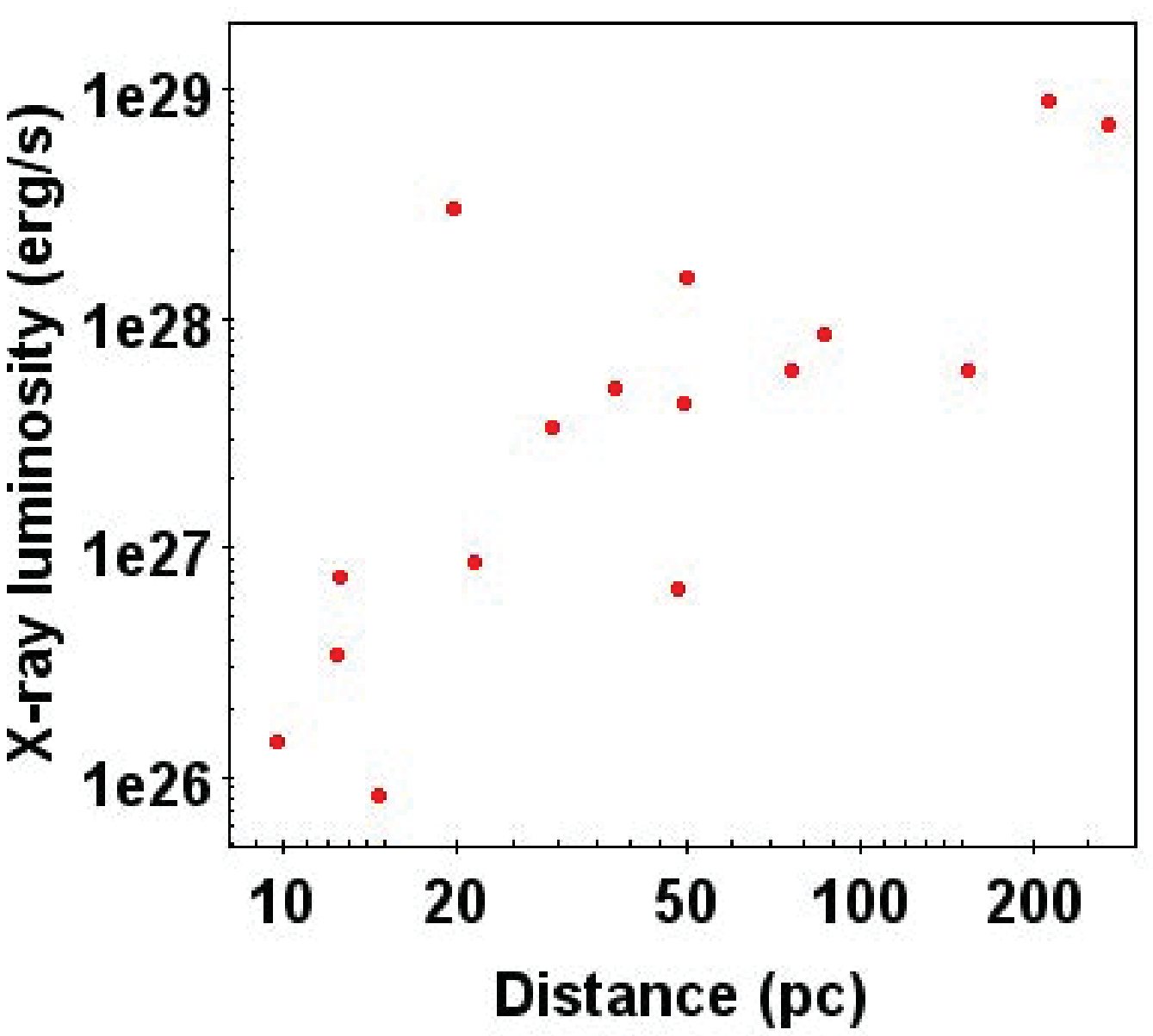}
  \caption{Examples of host-star properties, plotted from the catalogue content. }
   \label{fig_examples}
\end{center}
\end{figure}

Fig.\ref{fig_examples} illustrates a few of the stellar parameters from our initial assessments to date.

In order to fill in gaps in the stellar parameter space (e.g.\ to have adequate UV and X-ray measurements across spectral types) we are planning observing proposals, principally to HST and XMM-Newton.

\section{Determination of the physical parameters of the host stars}

\begin{figure}[h]
\begin{center}
 \includegraphics[width=65mm]{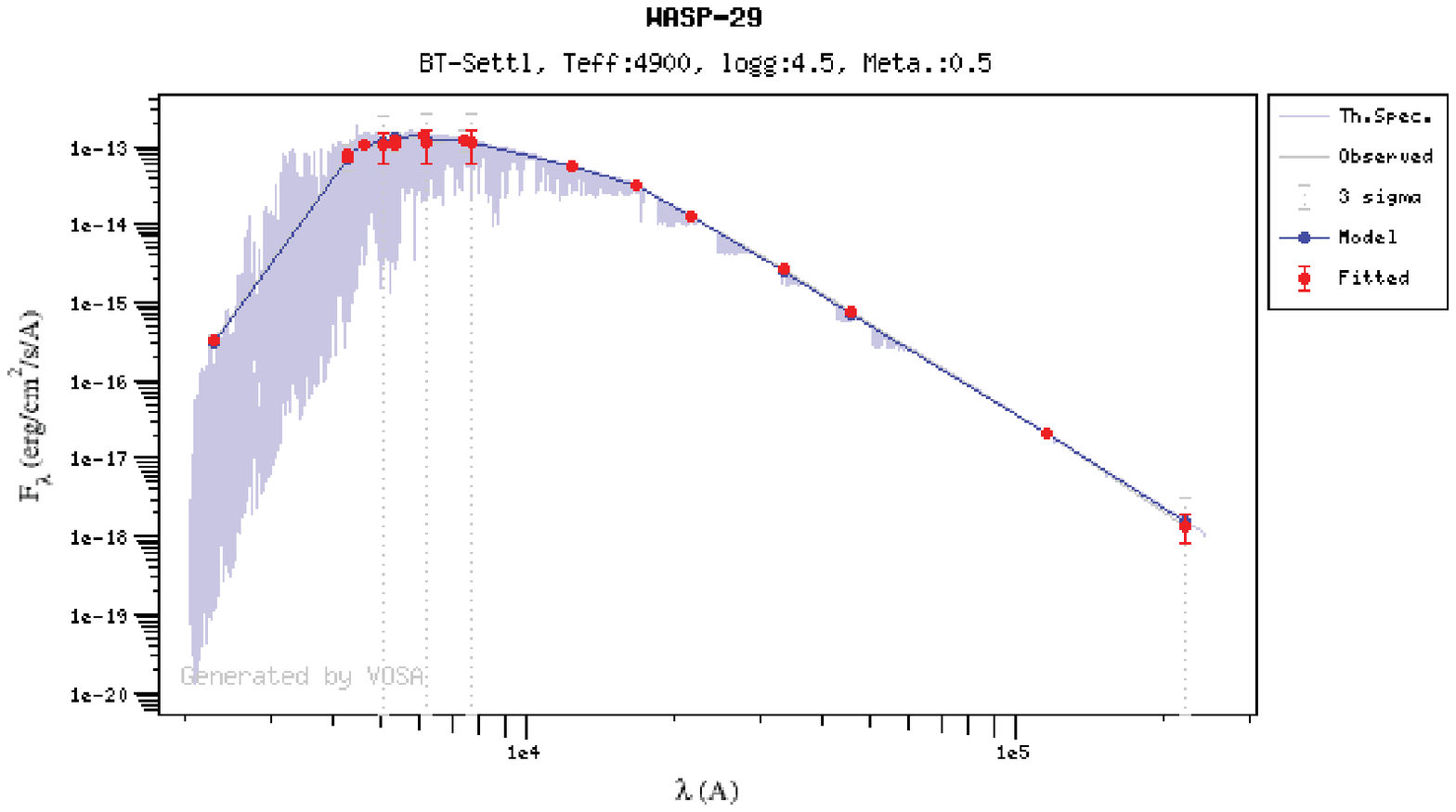}
 \includegraphics[width=65mm]{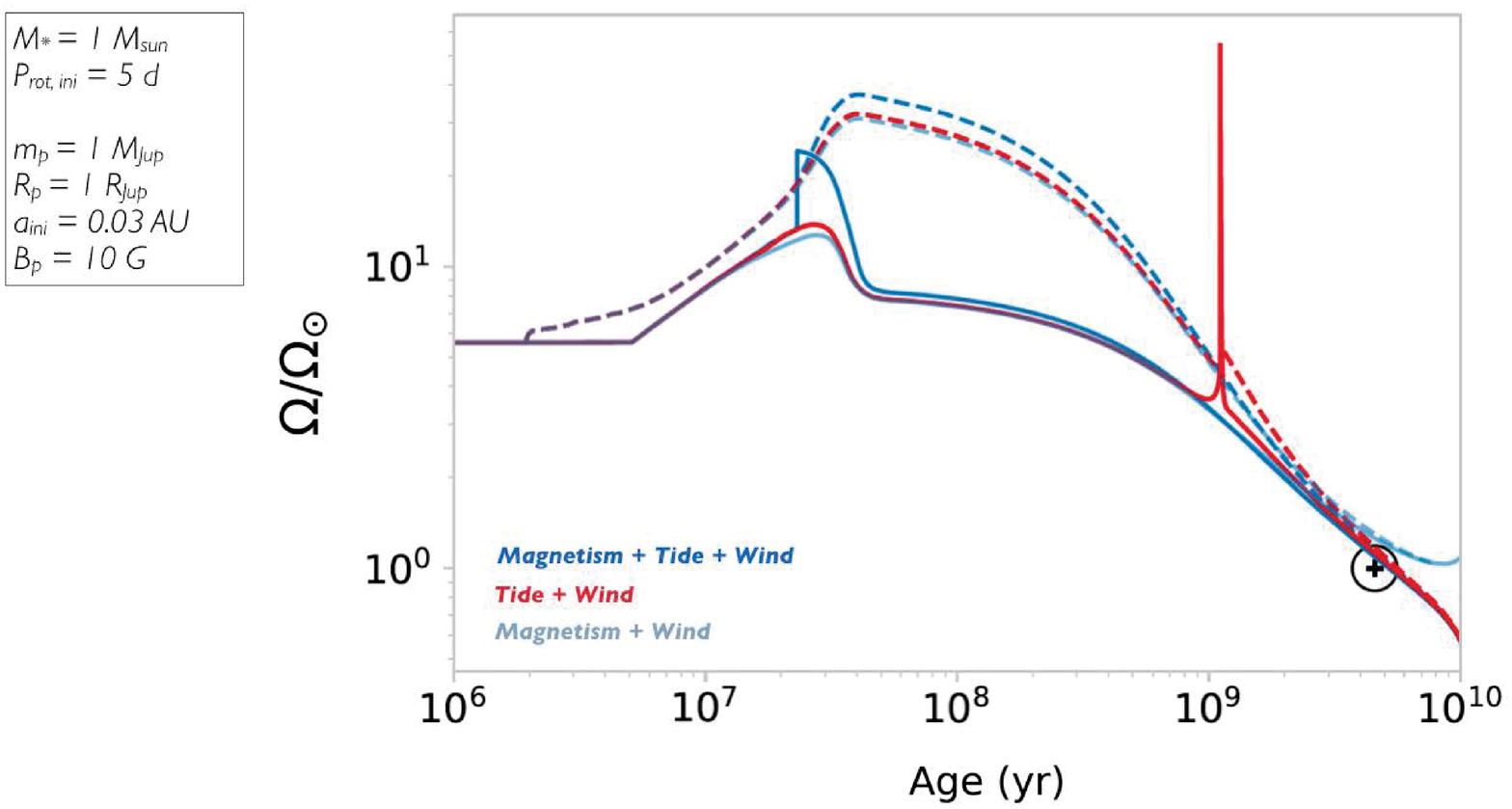}
  \caption{{\it (a, left).} SED fitting using VOSA. Blue spectrum represents the theoretical model (Allard \etal\ 2012) that best fits while red dots represent the observed photometry. {\it (b, right).} Stellar rotation rate (days) as a function of log(age, years) for 3 cases: wind + Alf\'en wings torques (light blue curve), wind + tides (red) and wind + tides + Alfv\'en wings torques (dark blue) (Ahuir \etal\ 2019 in prep; see Benbakoura \etal\ 2018 for details of the ESPEM code used for the study). }
   \label{fig_sed_rot}
\end{center}
\end{figure}

A coherent and uniform determination of the stellar properties is essential to avoid any bias in the final results; use of Virtual Observatory (VO) tools facilitates this goal. 
For example, we have used VOSA (http://svo2.cab.inta-csic.es/theory/vosa/) to build observational Spectral Energy Distributions (SED) and  compare them with theoretical SEDs (Fig.\ref{fig_sed_rot}) to obtain physical parameters such as effective temperature, 
stellar radius and luminosity. 
Effective temperatures range from 3000 to 6500K. 
We have also used VOSA to search for infrared excess in our sources, with negative outcome. 

Medium/high resolution spectra for 48 of our sources 
observed with FEROS, HARPS, FORS 1-2, X-SHOOTER, FLAMES and UVES were gathered from the ESO archive. 
Visible-light photometry from Kepler/K2 and TESS will also be used to help characterise the stars in terms of rotational and aperiodic variability.
One third of our sources have 
Kepler light-curves while TESS will probably provide data for the remainder.

The ExoplANETS-A database of stellar properties and spectral models will be made publicly available, via a web-based interface, by the end of 2020.

\section{Scientific interpretation of the data: star--planet interactions}
Most exoplanets live around active stars, whose magnetism and intense activity can have
a direct impact on habitability conditions and more generally on the exosphere of the planet.
Intense wind and storms can lead to atmospheric loss. Further, a large fraction of exoplanets live
close to their host star (within 10--20 solar radii), to a point where in many cases they orbit within the star's Alfv\'en surface.
This has direct consequence on star--planet interaction and planet migration, as magnetic torques through
Alfv\'en wings directly connecting  the planet to its host star can occur. We have developed both
ab-initio 3D MHD simulation of such close-in systems (Strugarek \etal\ 2015, Strugarek \etal\ 2017) as well as a simpler 1D secular evolution model
of star--planet systems (Benbakoura \etal\ 2018). 
In Fig.\ref{fig_sed_rot}(b) we show the evolution of a star--planet system subject to intense
magnetic torques through direct Alfv\'en wings connection (on top of tides and wind effect) and compare it to the case with magnetic torques and stellar wind only (light blue curve)
or tides + wind (red curve). 
We note that adding all the effects leads to a quick demise of the planet, that impacts the star's rotation rate.
Such Alfv\'en wings can also lead to hot spots on the star's surface, impacting directly their light-curve and hence the transit curves.
We intend to develop 3D and secular models for the most important systems identified in the ExoplANETS-A project,
as well as obtaining spectropolarimetric magnetic maps of the host star in order to model the stellar wind and assess as accurately as possible the
space environment around these exoplanetary systems.

\end{document}